\documentclass{aip-cp}

\usepackage[numbers]{natbib}
\usepackage{rotating}
\usepackage{graphicx}
\usepackage{slashed}

\begin{document}

% Title portion
\title{Implications of Mixed Axion-Neutralino Dark Matter}

\author[aff1]{Hasan Serce\corref{cor1}}

\affil[aff1]{Department of Physics and Astronomy, University of Oklahoma, Norman, OK 73019, USA}
\corresp[cor1]{serce@ou.edu}

\maketitle

\begin{abstract}
The lack of evidence for weak scale supersymmetry from LHC Run-I and Run-II results along with null results from direct/indirect dark 
matter detection experiment have caused a paradigm shift in expected phenomenology of SUSY models. The SUSY dark matter candidate, neutralino $\widetilde{Z}_1$, 
can only satisfy the measured dark matter abundance due to resonance- and co-annihilations in cMSSM model. Moreover, viable parameter space is highly fine-tuned in the cMSSM. 
In models that can still satisfy the {\it naturalness} condition such as NUHM2, the neutralino is underproduced due to its higgsino nature. Neutralino 
combined with axion that solves the strong CP problem can explain the observed dark matter in the universe. Here I briefly discuss implications 
of the mixed axion-neutralino scenario.
\end{abstract}

% Head 1
\section{Introduction}

The discovery of the Higgs boson with a mass $m_h = 125.09 \pm 0.21$ (stat.) $\pm  0.11$ (syst.) GeV~\cite{Aad:2012tfa,Chatrchyan:2012xdj} confirms the particle 
content of the Standard Model. Although the measured value of the Higgs boson squarely fits in the region 75-135 GeV as predicted by the 
minimal supersymmetric standard model (MSSM)~\cite{Carena:2002es}, lack of signal for sparticles raises questions on naturalness of the supersymmetric (SUSY) models. 
The SUSY spectrum was 
predicted to lie not too far from the weak scale $\sim 100$ GeV, based on {\it naturalness} calculations oftenly expressed via BG measure~\cite{Barbieri:1987fn}:
\begin{equation}
 \Delta_{\rm BG}\equiv \max \left[ c_i\right]\ \ {\rm where}\ \ 
c_i=\left|\frac{\partial\ln m_Z^2}{\partial\ln p_i}\right|
=\left|\frac{p_i}{m_Z^2}\frac{\partial m_Z^2}{\partial p_i}\right|
\label{eq:DBG}
\end{equation}
where $p_i$'s are the various parameters of particular effective theories. $\Delta_{\rm BG}$ measures how sensitive the $Z$-boson mass is to variations of 
parameters at some high defining scale. In such a measure the gluino mass, lower bound of which is set to $\simeq1.9$ TeV by ATLAS group~\cite{ATLAS:2016uzr}, was expected 
to be less than 350 GeV. 
Since LHC searches pushed sparticle (SUSY particle) masses 
to multi-TeV energy scale, remaining supersymmetric models are considered to be in crisis.
It has been argued that $\Delta_{\rm BG}$ overestimates fine-tuning 
when applied to effective theories with multiple independent soft terms that are correlated~\cite{Baer:2013gva,Mustafayev:2014lqa}. 
The electroweak fine-tuning~\cite{Baer:2012up}, 
$\Delta_{\rm EW}$, 
is a model independent fine-tuning measure which compares the largest contribution on the right-hand side of Eq.~(\ref{eq:mzs}) 
to the value of $m_Z^2/2$:
\begin{equation}
 \frac{m_Z^2}{2} = \frac{(m^2_{H_d}+\Sigma_d^d)-(m^2_{H_u}+\Sigma_u^u)\tan^2\beta}{(\tan^2\beta -1)}
-\mu^2
\label{eq:mzs}
\end{equation}
Eq.~(\ref{eq:mzs}) is the well-known condition from minimization of the Higgs potential for electroweak symmetry breaking to occur. The electroweak fine-tuning 
is defined as:
\begin{equation}
\Delta_{\rm EW} \equiv max_i \left(|C_i|\right)/(m_Z^2/2) 
\label{eq:deltaew}
\end{equation}
where $C_{H_u}=-m_{H_u}^2\tan^2\beta /(\tan^2\beta -1)$,
$C_{H_d}=m_{H_d}^2/(\tan^2\beta -1)$ and $C_\mu =-\mu^2$, along with definitions for the radiative corrections $C_{\Sigma_u^u(k)}$ and 
$C_{\Sigma_d^d(k)}$~\cite{Baer:2012cf}. Low $\Delta_{\rm EW}$ assures that there are no large cancellations on the right-hand side of Eq.~(\ref{eq:mzs}). 
Models with $\Delta_{\rm EW}<30$, which corresponds to 3\% or less fine-tuning, are considered as {\it natural}. This should not be considered as an 
attempt to save SUSY but an appropriate definition of naturalness to avoid overestimation for the models with multiple uncorrelated soft terms. From Eq.~(\ref{eq:mzs}), using   
the naturalness bound $\Delta_{\rm EW}<30$, the $\mu$ term is restricted to be less than 355 GeV. This is not a problem in models with non-unified Higgs masses since 
a weak scale value of $\mu$ can easily be chosen or $m_{H_u}^2$(GUT) can be adjusted so that it barely runs negative after RGE running. 

The constrained MSSM (cMSSM) model with only 4 parameters has been severely constrained by dark matter and sparticle searches. cMSSM models with 1 TeV higgsino are still 
viable~\cite{Roszkowski:2014wqa,Bagnaschi:2015eha} by giving up the naturalness constraint. 
Even before LHC was turned on, 
LEP-II working group~\cite{Barate:2003sz} 
reported the lower limit of Higgs mass as $m_h \gtrsim 114.4$ GeV in 2003, which forced natural cMSSM to survive in conflict with 
naturalness expectations~\cite{Barbieri:1998uv}. The 'WIMP miracle' picture with a bino-like neutralino had already been disfavored in MSSM~\cite{Baer:2010wm}
before the discovery of the Higgs boson.
In radiatively-driven natural SUSY models, neutralinos are underproduced due to higgsino-like neutralino so 
a second dark matter component ({\it e.g.}, axions) is needed. By introducing axion, another fine-tuning problem, namely the strong CP problem, is addressed.

\section{Neutralino LSP in Natural SUSY}
\begin{figure}[h]
  \centering
  \begin{tabular}[b]{c}
    \includegraphics[width=.46\linewidth]{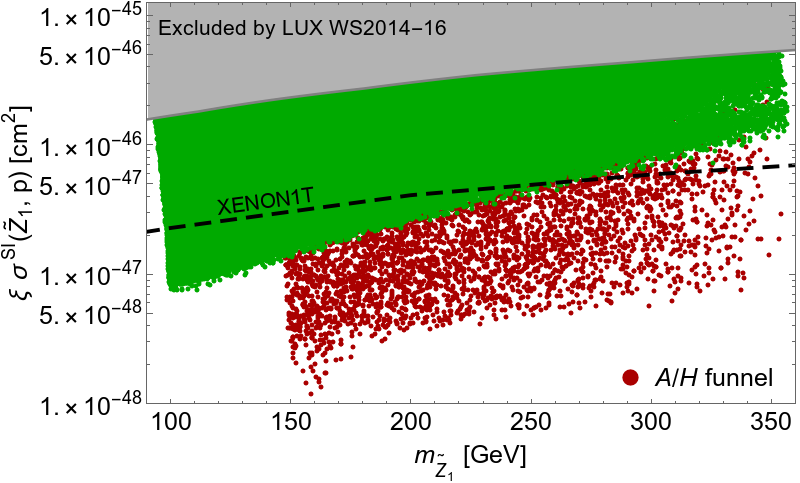} \\
    \small (a)
  \end{tabular} \qquad
  \begin{tabular}[b]{c}
    \includegraphics[width=.45\linewidth]{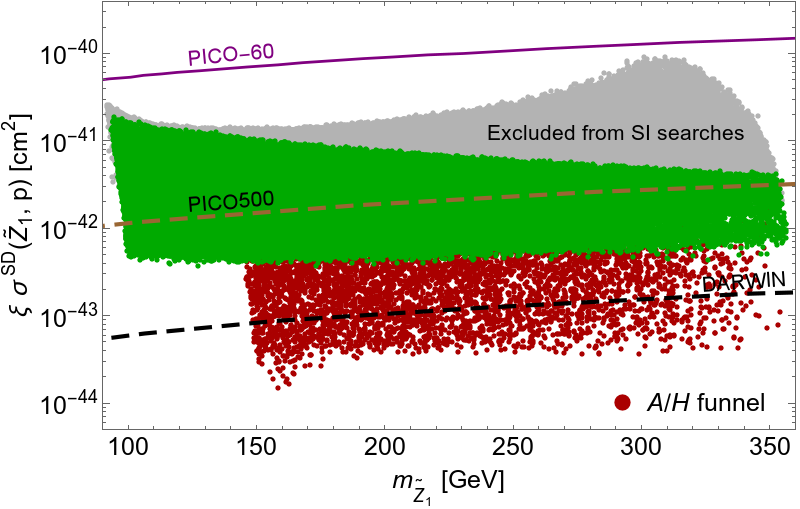} \\
    \small (b)
  \end{tabular}
  \caption{Plot of rescaled (a) spin-independent $\xi \sigma^{\rm SI}(\widetilde{Z}_1, p)$ and (b) spin-dependent 
  $\xi \sigma^{\rm SD}(\widetilde{Z}_1, p)$ direct detection rate  
vs. neutralino LSP mass $m_{\widetilde{Z}_1}$. All points satisfy the naturalness constraint: $\Delta_{\rm EW}<30$. Red points show the additional region that appears 
when $m_{A/H} \simeq 2 \times m_{\widetilde{Z}_1}$. Published DM search results are shown by straight lines. Future projected reaches are shown by dashed lines.}
\label{det}
\end{figure}

In natural SUSY models with $\Delta_{\rm EW}<30$, the lightest supersymmetric particle (LSP) is higgsino-like with a mass $m_{\widetilde{Z}_1} \simeq \mu$ so the 
neutralino mass is bounded above by the value of 
the $\mu$ parameter. In such a model, a pure higgsino-like neutralino can 
only make up one fourth of the total dark matter abundance considering thermal production only. Thermally produced WIMP abundance can be higher with a considerable 
amount of bino mixing by considering low $m_{1/2}$ but such a parameter 
set has already been ruled out by the LUX dark matter and LHC gluino searches unless $\tan\beta$ is small~\cite{Badziak:2017the}. 
Spin-independent (SI) and spin-dependent (SD) WIMP-proton cross sections for the NUHM2 model with $\Delta_{\rm EW}<30$, calculated using ISAJET v7.86~\cite{Paige:2003mg}, 
is shown in Fig.\ref{det}.

Scattering cross section rates are scaled down by a factor $\xi=\Omega_{\widetilde{Z}_1}^{th}h^2/0.12$~\cite{Bottino:2000jx} due to the fact that the WIMPs comprise only
a portion of the local dark matter abundance. The higher bound is set by LUX bounds on the spin-independent scattering cross section~\cite{Akerib:2016vxi} 
whereas the lower bound is determined by the naturalness condition.
Unlike in the cMSSM, $A/H$ funnel might occur for any $\tan\beta$ values since $m_A$ is an input 
parameter in the NUHM2 model given that $m_A$ vs. $\tan\beta$ bounds~\cite{Aaboud:2016cre} are respected and shown by red dots in Fig.\ref{det}. The upper bound on 
the BR$(b \to s \gamma)$ removes the parameter region where $m_{A/H} \lesssim 300$ GeV~\cite{Bae:2015nva}.
Green points represent the region where 
no additional annihilation mechanism is required~\cite{Baer:2016ucr}. It is questionable to expect $A/H$ resonance, 
or why the nature chooses 
$m_{A/H} \simeq 2 \times m_{\widetilde{Z}_1}$ in the SUSY 
scenarios with non-unified scalar masses. Neutralino abundance becomes lower with extra annihilations so $\xi$ gets even smaller, which results in a lower detection rate due 
to fewer target particles. Stau co-annihilation region does not appear as an additional region since $m_{\widetilde{Z}_1} \lesssim 360$ GeV 
and the lower bound for $m_{1/2}$ is rather high from LHC searches for a light stau mass. Recent results from PICO-60 collaboration~\cite{Amole:2017dex} 
are shown in Fig.\ref{det}(b). Projected spin-dependent dark matter searches such as DARWIN~\cite{Aalbers:2016jon} are not able to probe the 
whole parameter space predicted by the natural NUHM2 model. The gray region shows already excluded 
points from SI LUX searches. 

On the SI-plane, XENON1T dark matter search~\cite{Aprile:2015uzo} will be able to cover the bulk of the remaining parameter space. 
Projected future reaches from n-tonne detectors such as LZ~\cite{Akerib:2015cja}, XENONnT~\cite{Aprile:2015uzo}, 
DarkSide-20K~\cite{Agnes:2015lwe}, DEAP-50T~\cite{Amaudruz:2014nsa} and DARWIN~\cite{Aalbers:2016jon} will cover the remaining parameter space of the NUHM2 model.

In natural generalized mirage mediation model (nGMM)~\cite{Baer:2016hfa}, wino and bino masses may be elevated that result in a lower rescaled scattering rate. 
Although XENON1T will not be able to cover most of the parameter space, n-tonne detectors will probe the whole parameter space.

\section{Supersymmetrized DFSZ axion model}

The QCD Lagrangian contains a CP-violating term:
\begin{equation}
 {\cal L}\ni \bar{\theta}\frac{g_s^2}{32\pi}G_{A\mu\nu}\tilde{G}^{\mu\nu}_A
\end{equation}
where $\bar{\theta}\equiv\theta+arg\ det (M)$ and $M$ is the quark mass matrix. Measurements of the neutron electric dipole moment (EDM) imply that 
$\bar{\theta}\ll 10^{-10}$ thus requiring a huge fine-tuning in $\bar{\theta}$. The smallness of $\bar{\theta}$ is known as the strong CP problem. 
A solution to the problem is to introduce Peccei-Quinn (PQ) symmetry~\cite{Peccei:1977hh} which causes the $G\tilde{G}$ term dynamically to settle
to zero when $U(1)_{\rm PQ}$ is broken. The associated pseudo Nambu-Goldstone boson with the PQ symmetry breaking is called 
{\it axion}, $a$~\cite{Weinberg:1977ma,Wilczek:1977pj}.

In the supersymmetrized DFSZ scenario, MSSM lagrangian is augmented by:
\begin{equation}
 W_{\rm DFSZ}\ni \lambda\frac{S^2}{M_{Pl}}H_uH_d 
\label{eq:KN}
\end{equation}
where $S$ is a singlet superfield charged under PQ symmetry. Higgs doublet {\it superfields} $H_u$ and $H_d$ carry PQ charges so the SUSY $\mu$ term is in fact 
forbidden before the PQ symmetry-breaking~\cite{Kim:1983dt}. An effective $\mu$ term is generated with:
\begin{equation}
 \mu\sim \lambda \mbox{ } v_{\rm PQ}^2/M_{Pl}. 
\end{equation}
A weak scale $\mu$ term can easily be generated by breaking PQ symmetry radiatively~\cite{Murayama:1992dj}. Then little hierarchy characterized by 
$\mu\sim m_Z\ll m_{3/2}\sim {\rm multi-TeV}$ emerges quite naturally due to the mis-match between PQ breaking scale and hidden sector 
mass scale $f_a\ll m_{\rm hidden}$.
In a Peccei-Quinn augmented MSSM (PQMSSM) scenario, the axion superfield is given by~\cite{Bae:2011jb}:
\begin{equation}
  A = \frac{1}{\sqrt{2}}\left(s+ia\right) + \sqrt{2}\theta \tilde a + \theta^2 F_a
\end{equation}
where $a$ is the axion field, $s$ is the spin-0 {\it saxion} field and $\tilde a$ is spin-$\frac{1}{2}$ fermionic partner of axion called {\it axino}. 
In addition to the thermal production, in PQ-augmented SUSY scenarios WIMPs are produced by subsequent decay of both axino ($\tilde{a}\to \widetilde{Z}_1+...$) 
and saxion ($s\to \widetilde{Z}_1\widetilde{Z}_i$) when kinematically allowed. 
The $s \to aa / \tilde{a}\tilde{a}$ branching ratio 
is controlled by the axion-saxion effective coupling~\cite{Chun:1995hc}:
\begin{equation}
\mathcal{L} \ni \frac{\xi_s}{f_a} s \left[ \left(\partial_\mu a \right)^2 + i \bar{\tilde{a}} \slashed\partial \tilde{a} \right]
\label{eq:xis}
\end{equation}
where $\xi_s$ can take any values between 0 and 1. Although saxion decays to axion pairs at large axion decay constant, $f_a$, can 
produce significant amount of dark radiation, dark matter density constraint is always the most restrictive one for a saxion mass at TeV scale.

Total neutralino abundance can be computed accounting for adding thermally produced neutralino and neutralino production from decays of axinos and saxions. 
Amount of axion needed 
to satisfy the measured DM abundance~\cite{Calabrese:2017ypx}, 
$\Omega_{a}^{\rm co} h^2=0.12-\Omega_{\widetilde{Z}_1}^{th}h^2$, is produced from the coherent oscillations of the axion field. Desired amount of axion can be computed 
by adjusting the misalignment angle $\theta_i$~\cite{Visinelli:2009zm}:
\begin{equation}
 \Omega_a^{\rm co} h^2 \simeq 0.23
f(\theta_i)\theta_i^2\left(\frac{f_a}{10^{12} \mbox{ GeV}}\right)^{7/6}
\label{eq:axco} 
\end{equation}
where $f(\theta_i)$ is the anharmonicity factor, parametrized as  
$f(\theta_i)= \left[\ln\left(e/(1-\theta_i^2/\pi^2\right)\right]^{7/6}$. The axion misalignment angle $\theta_i$ can take any values between 
$-\pi$ and $\pi$. However, $f(\theta_i)$ is very sensitive to a small change in $\theta_i$ for $\theta_i \gtrsim 3$ hence 
the parameter set that satisfy $\Omega_a^{\rm co} h^2 + \Omega_{\widetilde{Z}_1}h^2=0.12$ for $\theta_i >\pi$ is considered as {\it unnatural}. In SUSY DFSZ model, 
this region occurs when the axion decay constant, $f_a \lesssim 10^{11}$ GeV~\cite{Bae:2014rfa,Bae:2015rra}. PQ breaking is assumed to have occured before or during 
inflation and has not been restored, to avoid any domain wall problem.

Axion-neutralino mixed dark matter can be calculated by solving the eight coupled Boltzmann equations~\cite{Bae:2014rfa}. Thermal production of 
axion, axino and saxion are independent of the reheat temperature, $T_R$, in the SUSY DFSZ scenario so the gravitino problem is avoided for the values of 
$T_R$ less than $\sim 10^{10}$ GeV~\cite{Bae:2015efa}. The evolution of number densities of gravitino, neutralino, axino and saxion 
(both thermally and coherently produced) are tracked from the end of inflation, from $T_R$, until today. Lifetimes of axino, gravitino and saxion are tracked along 
with their abunances in order to check whether there is a violation to big bang nucleosynthesis (BBN) or not. Although saxions mainly decay into axion pairs, 
in cases where saxions 
are light $m_s \ll m_0$, long-lived saxions might impose stronger constraint from BBN on the maximum value of $f_a$ than DM abundance constraint. Nevertheless, in 
models with gravity mediated SUSY breaking saxion mass is expected to be at the order of gravitino mass 
$m_s \simeq \alpha m_{\tilde{G}}$. For each parameter set which yields $\Omega_{\widetilde{Z}_1}h^2<0.12$, the axion misalignment angle $\theta_i$ is adjusted using 
Eq.~(\ref{eq:axco}) so that 
$\Omega_{\widetilde{Z}_1}h^2+\Omega_a^{\rm co} h^2=0.12$.

\begin{figure}[h]
  \centerline{\includegraphics[width=350pt]{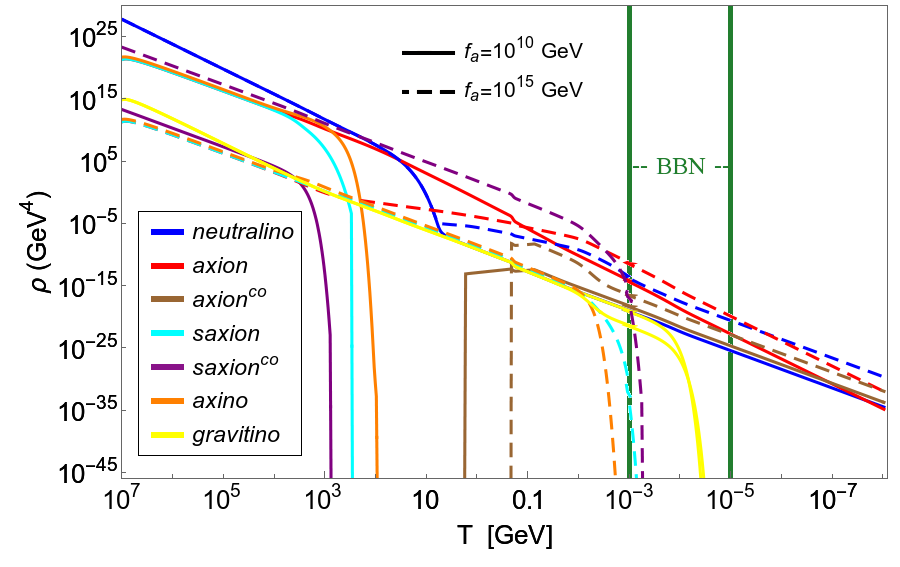}}
  \caption{Evolution of the energy densities of axion, axino, gravitino, neutralino and saxion vs. temperature, $T$ from the reheat temperature 
  $T_R=10^7$ GeV to now. Two scenarios with different axion decay constants for $f_a=10^{10}$ GeV (solid) 
  and $f_a=10^{15}$ GeV (dashed) are illustrated. Vertical green lines bound the temperature range where BBN takes place.}
\label{evols}
\end{figure}

An example of evolution of the energy densities from the solution of coupled Boltzmann equations 
is illustrated in Fig.\ref{evols} for the same NUHM2 benchmark point (parameters given in the next section) with different $f_a$ values. 
Gravitino and axino masses are 
set to 10 TeV whereas saxion mass is set to 5 (10) TeV for the large (small) $f_a$ case to show a scenario where BBN violation occurs. 
The reheat temperature is chosen at $10^7$ GeV. Solid lines show the evolutions for $f_a=10^{10}$ GeV. The point is safe from BBN constraint and 
$\Omega_{\widetilde{Z}_1}h^2 = \Omega_{\widetilde{Z}_1}^{th}h^2 \simeq 0.006$ since axinos and saxions decay significantly before the neutralino freeze-out. As a result, 
the cold dark matter density is mainly from coherent production of axions with $\theta_i \simeq \pi$. Even though the gravitino decays during BBN, its abundance is not 
large enough to intervene the nucleosynthesis.
Dashed lines show the evolutions for $f_a=10^{15}$ GeV; this point is not allowed from BBN and dark matter density constraints. Saxion$^{\rm co}$s are still decaying when 
nucleosynthesis starts at $\sim1$ MeV so the point is not BBN-safe. For a larger $f_a$, thermal yields 
of axinos, axions and saxions at $T=T_R$ are lower but their couplings to matter is weaker hence they decay much later. 
The axions from $s\to a +a$ decays can be seen 
as a rise in the (relativistic) axion component.
In both cases, axion$^{\rm co}$ production starts at $T\sim 1$ GeV. For the $f_a=10^{15}$ GeV case, dark matter is already overproduced so the axion misalignment 
angle is set to 1 for simplicity. Scan results for the same benchmark point with $\xi_s=0$ gives an upper bound on $f_a=2 \times 10^{12}$ GeV.

\section{Admixture of neutralino-axion in SUSY DFSZ}
In the SUSY DFSZ scenario, thermal productions of axinos and saxions are proportional to $1/f_a^2$ whereas coherent production of axions 
and saxions increase with increasing $f_a$. 
In the lower $f_a$ region 
$10^9 \leq f_a/$GeV $\leq 10^{13}$ where the lower bound is from astrophysical observations, axino decays mainly contribute to neutralino abundance. In the region 
$10^{13} \leq f_a/$GeV $\leq 10^{16}$, direct or indirect decays of saxion dominantly augment the neutralino abundance. Although yields of thermally produced 
axinos and saxions decrease with increasing $f_a$, their lifetime increase since their couplings become weaker. The amount of allowed neutralino density from 
the decays is constrained by DM searches.

\begin{figure}[h]
  \centerline{\includegraphics[width=350pt]{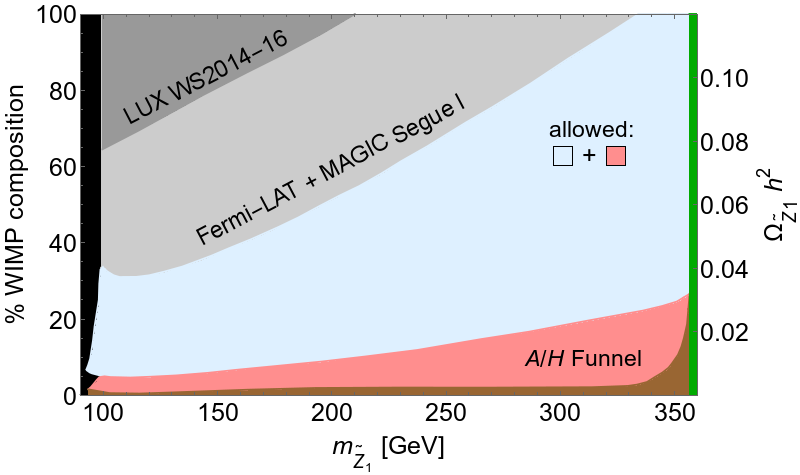}}
  \caption{Percentage composition of higgsino-like neutralino allowed in neutralino-axion admixture from NUHM2 model. Red region is  
  viable due to enhanced annihilations. Black and gray shaded regions are excluded by LEP-II searches and indirect (and direct, darker gray) DM searches respectively. 
  Green region is not allowed by the fine-tuning constraint. Brown shaded region is not viable due to very low thermal production of the neutralino LSP.}
\label{compos}
\end{figure}

In Fig.\ref{compos}, the amount of allowed neutralino dark matter in a 2-component DM scenario is shown quantitatively. For each point shown in Fig.\ref{det}, 
the maximum allowed $\xi$ ratio is computed without violating the LUX bound on the SI scattering cross section rate and the Fermi-LAT/MAGIC combined reach 
via on gamma rays from  $\widetilde{Z}_1 \widetilde{Z}_1 \to W^+W^-$ channel~\cite{Ahnen:2016qkx}. Although indirect dark matter searches have not started probing the region expected from the NUHM2 model as seen in Fig.3 
of Ref.\cite{Baer:2016ucr}, results from 
Fermi-LAT/MAGIC collaboration put a stronger constraint on the allowed neutralino abundance since the annihilation rate is rescaled by $\xi^2$. 
The blue and pink regions are the allowed regions for the amount of neutralinos present in the admixture. 
The lower edge of the blue region can be read as the maximum amount of neutralinos produced thermally in the natural NUHM2 model. Due to the higgsino-like neutralino, 
neutralinos 
can only make up to $\simeq 25$\% of the dark matter abundance. Its composition can be augmented up to any allowed point in the blue region.
Additional neutralinos are assumed to be produced from axino and saxion decays. 
For $m_{\widetilde{Z}_1} \gtrsim 330$ GeV, neutralino 
can make up to 100\% of the DM without violating published limits on the dark matter annihilation cross section rate~\cite{Ahnen:2016qkx}.
%However, upper bound of the blue region is strictly constrained by the LUX searches and calculated cross section scattering rate is just below the LUX exclusion limit. 
Thermally produced neutralino abundance can be as low as $\sim$0.005 with $A/H$ resonance annihilations (pink shaded area). 
The black region is excluded from LEP-II searches, $m_{\widetilde{W}_1^{+/-}}>103.5$ GeV whereas in the gray shaded region Fermi-LAT/MAGIC exclusion applies. In the green area, 
electroweak fine-tuning is big, $\Delta_{\rm EW}>30$, so considered to be unnatural. In the brown shaded region, thermally produced neutralino abundance is too low to reach 
even with enhanced annihilations. 

\begin{figure}[h]
  \centering
  \begin{tabular}[b]{c}
    \includegraphics[width=.47\linewidth]{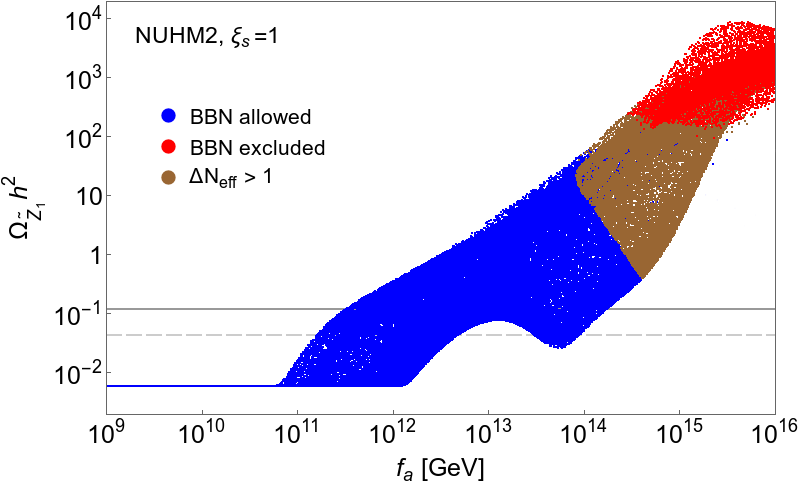} \\
    \small (a)
  \end{tabular} \qquad
  \begin{tabular}[b]{c}
    \includegraphics[width=.47\linewidth]{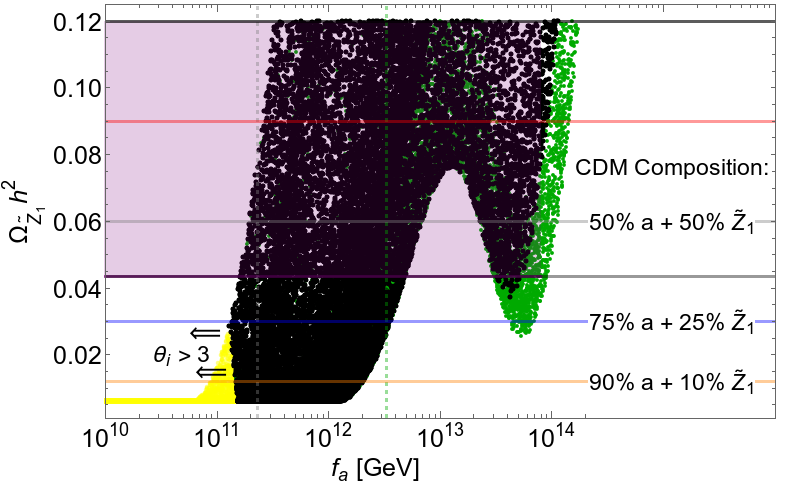} \\
    \small (b)
  \end{tabular}
  \caption{Plot of neutralino density vs. $f_a$ for the NUHM2 benchmark point. In frame (a), points that violate BBN and $\Delta_{N_{\rm eff}}$ constraints are colored 
  red and brown respectively. In frame (b), the region with $\Omega_{\widetilde{Z}_1}h^2 < 0.12$ is zoomed in. Purple shaded points are excluded from the indirect DM 
  searches. Green points are generated with both $m_{\tilde{a}}$ and $m_s$ greater than 30 TeV.}
\label{bolt}
\end{figure}

Neutralino cold dark matter density, 
$\Omega_{\widetilde{Z}_1}h^2=\Omega_{\widetilde{Z}_1}^{th}h^2+\Omega_{\widetilde{Z}_1}^{dec}h^2$, vs. axion decay constant, $f_a$, from 
a scan over $f_a:10^{9-16}$ GeV, $m_{\tilde{a}/s}:0.5-40$ TeV and $m_{\tilde{G}}=10$ TeV for 
a NUHM2 benchmark point with parameters:
\begin{equation}
 (m_0,\ m_{1/2},\ A_0,\ \tan\beta,\ \mu,\ m_A) = \mbox{(5300 GeV, 2030 GeV, -9850 GeV, 9, 150 GeV, 3000 GeV)}
\end{equation}
is shown in Fig.\ref{bolt}. The point has $\Delta_{\rm EW}=29.2$, $m_{\tilde{g}}=4481$ GeV within HE-LHC33 reach~\cite{Baer:2017yqq}, 
$\Omega_{\widetilde{Z}_1}^{th}h^2 \simeq 0.006$
and $<\sigma .v>\simeq4\times10^{-26}$ cm$^3$ s$^{-1}$. This is an example of a restricted NUHM2 model from indirect WIMP searches with $m_{\widetilde{Z}_1}\simeq 150$ GeV. 
General results from the scan with BBN and $\Delta_{N_{\rm  eff}}$ constraints are shown in 
Fig.\ref{bolt}(a). Points with $\Delta_{N_{\rm  eff}}>1$ are excluded at greater than 99\% confidence~\cite{Ade:2015xua} and colored brown. 
In Fig.\ref{bolt}(b), $f_a$ region that predicts $\Omega_{\widetilde{Z}_1}h^2 \leq 0.12$ is zoomed in.
Thermally produced neutralinos make only 5\% of the total dark matter density, its composition in the admixture 
can be augmented up to 36\% without violating Fermi-LAT/MAGIC combined reach via $W^+W^-$ channel. 
$\xi_s=1$ so $s \to aa$ and $s \to \tilde{a} \tilde{a}$ decay channels are open.

In the low $f_a$ region, $f_a \lesssim 2 \times 10^{10}$ GeV, axinos and saxions decay before the neutralino freeze-out so $\Omega_{\widetilde{Z}_1}h^2$ takes its 
standard thermal value $\Omega_{\widetilde{Z}_1}^{th}h^2$ which is independent of PQ parameters : $f_a$, $m_{\tilde{a}}$, $m_s$, $\theta_{i/s}$ and $\xi_s$.
Axinos and saxions decay more slowly with increasing $f_a$ since their couplings to particles / sparticles are proportional to $1/f_a$. Only long-lived 
axinos and saxions enhance neutralino DM density. Neutralino relic density strictly increases with increasing $f_a$ for $f_a \lesssim 10^{13}$ GeV. 
For $f_a \gtrsim 10^{13}$ GeV, neutralino density indeed decreases for the  points with $m_s \lesssim 2m_{\tilde{a}}$. 
%since $s \to aa / \tilde{a} \tilde{a}$ decays 
%dominate over $s \to \widetilde{Z}_1 +  \widetilde{Z}_1$ decay~\cite{Bae:2013bva}.
The saxion mainly decays to axion pairs and the decay $s\to \tilde{a} \tilde{a} \to \widetilde{Z}_1 + X$ is kinematically not allowed.  
Moreover, $s \to SM$ decay injects entropy into the universe that dilutes 
relics. In the high $f_a$ region, saxions are produced coherently in a large amount, hence their decays increase neutralino density even though 
BR$(s \to \widetilde{Z}_1 +  \widetilde{Z}_j)$ is suppressed.

For the benchmark point, BBN violation starts at $f_a \simeq 2 \times 10^{14}$ GeV and $\Delta_{N_{\rm  eff}} >1$ for $f_a \simeq 8 \times 10^{13}$ GeV. Nonetheless; 
such points have already been excluded from the dark matter density constraint. 
The most constraining upper bound for $f_a$ is from Fermi-LAT/MAGIC exclusion : $\Omega_{\widetilde{Z}_1}h^2 \leq 0.043$. 
The excluded region from the indirect DM search is shaded 
purple in Fig.\ref{bolt}(b). $f_a$ values greater than $2 \times 10^{14}$ GeV are not allowed. 
In the $f_a$ range between $6\times 10^{12}$ GeV and $3\times 10^{13}$ GeV, WIMPs are overproduced. 
For some parameter choices with a softer constraint from the indirect DM searches, 
%or $m_{\tilde{a}/s} \gtrsim 30$ TeV, 
the diluted region can be within the allowed range; a continous range of $f_a$ up to $2 \times 10^{14}$ GeV can be allowed. 
For saxion and axino masses less than $\sim$30 TeV (black dots), the upper bound on $f_a$ is $6 \times 10^{12}$ since entropy injection from saxion decays can not lower 
$\Omega_{\widetilde{Z}_1}h^2$ under the allowed region. 
The yellow points show the results with $\theta_i \gtrsim 3$. Considering naturalness in the PQ sector, 
values of $\theta_i \sim \pi$ are fine-tuned so the lower bound for $f_a$ in this natural DFSZ scenario is $10^{11}$ GeV. In the scenario shown by black dots only, 
the axion is expected to have a 
mass $2$ $\mu$eV $\lesssim m_a \lesssim$ $60$ $\mu$eV which is mostly in the range of the projected sensitivity by ADMX Gen2 experiment~\cite{Stern:2016bbw}. 
The large $f_a$ region which is accessible with heavy $m_{\tilde{a}/s}$ is not within the range of axion search experiments. 
In a simple scenario where the axion is the only dark matter candidate, setting $\theta_i=1$, the axion is overproduced if its mass is lighter than $\sim 2$ $\mu$eV. 
ADMX Gen2 has search 
capabilities for axion mass from 2 to 40 $\mu$eV with a high DFSZ sensitivity up to $m_a \simeq 25$ $\mu$eV.

For the $\xi_s=0$ case, in which the decays $s\to \tilde{a} \tilde{a}$ and $s\to a a$ are turned off, the upper bound on $f_a$ is more severe. In this case, 
$\Omega_{\widetilde{Z}_1}h^2$ is strictly increasing with increasing $f_a$ since BR$(s \to \tilde{a} +  \tilde{a}) =0$ and there is no dilution mechanism 
from $s\to SM$ injection to the thermal bath.

\section{Summary}
The natural SUSY scenario is being probed by LHC searches and WIMP searches. 
Projected n-tonne direct WIMP detection experiments (DarkSide, DEAP, LZ, XENONnT, DARWIN) reaches cover almost the entire motivated SUSY dark matter models. 
Moreover, DARWIN is projected to detect very low scattering cross section, almost down to the neutrino background, for WIMP masses between 0.1-1 TeV. WIMPs in natural SUSY  
are expected to have a mass less than $\sim$350 GeV, which is a unique signature for underabundant SUSY DM scenarios. In such models, WIMP 
(higgsino-like neutralino) detection is ultimately expected. As $g_{a\gamma \gamma}$ coupling in SUSY DFSZ model is lower than the one in non-SUSY DFSZ axion model, 
it is not clear whether ADMX Gen2 searches will be able to reach SUSY DFSZ predicted $g_{a\gamma \gamma}$ coupling or not. An axion detection within projected 
reaches is an indication of mainly axion admixture where neutralino contribution to total DM density is low. 

% Acknowledgement
\section{Acknowledgements}
I would like to thank K.J. Bae, H. Baer, V. Barger and A. Lessa for earlier collaborations on this topic and Caroline Serce for proof-reading. 
The computing for this project was performed at the OU Supercomputing Center for Education \& Research (OSCER) at the
University of Oklahoma (OU).
% References

\nocite{*}
\bibliographystyle{aipnum-cp}%
\bibliography{natsusy}%

\end{document}